\documentclass[
 reprint,=
 amsmath,amssymb,
 aps,unsortedaddress]{revtex4-1}
\usepackage{xcolor}
\usepackage{graphicx}
\usepackage{dcolumn}
\usepackage{bm}
\usepackage{amsmath}
\usepackage{lipsum}
\usepackage{tikzsymbols}
\usepackage{mathtools}

\newcommand{\be}{\begin{equation}}
\newcommand{\ee}{\end{equation}}

\begin{document}

\title{Non-Abelian defects in Fracton phase of matter}

\author{Yizhi You}
\affiliation{\mbox{Princeton Center for Theoretical Science, Princeton University, Princeton NJ 08544, USA}}

\date{\today}
\begin{abstract}
Fracton phase of matter host fractionalized topological quasiparticles with restricted mobility. A wide variety of fracton models with abelian excitations had been proposed and extensively studied while the candidates for non-Abelian fracton phases are less explored. In this work, we investigate the effect of twisted defect in abelian fracton models. The twisted defect is launched by introducing a branch cut line hosting anyon condensate. In particular, these twisted defects, which alter different types of quasiparticles, carry projective non-Abelian zero modes. En route, such defects can be engineered via strong onsite hybridization along a branch cut which provides wide tunability and flexibility in experiment platforms. The braiding of twisted defects with projective non-Abelian Berry phase renders a new avenue toward fault-tolerant quantum computation.

\end{abstract}

\maketitle

\section{Introduction}
Strongly correlated system manifest a rich class of unconventional quantum matter with long-range entanglement and fractionalization. Such long-range entangled phase of matter, so-called topological phases, host quasi-particles carrying fractional quantum numbers and obey anyonic statistics\cite{wen1990topological,willett1987observation,wen1992classification,dijkgraaf1990topological}.

Recently, a new type of long-range entangled state in 3D-- fracton order---which has precursors in the study of glassiness\cite{Chamon2005-fc,castelnovo2012topological}, spin liquids\cite{yoshida2013exotic}, and quantum error correcting codes\cite{Haah2011-ny}, has drawn increasing attention\cite{Vijay2016-dr,Devakul2017-gg,Slagle2017-ne,Ma2017-qq,Halasz2017-ov,Hsieh2017-sc,Vijay2017-ey,Slagle2017-gk,Slagle2017-st,Williamson2016-lv,Ma2017-cb}. The most notable example of this phenomenon is the immobility and subdimensional nature of the “fracton” excitations\cite{pretko2017fracton,ma2018fracton,bulmash2018higgs,Slagle2017-gk,Slagle2017-la,shirley2017fracton,prem2018pinch,pretko2017fracton,Pretko2017-ej,slagle2018symmetric,gromov2017fractional,2018arXiv180711479P,pai2018fractonic,2018arXiv180601855B,pretko2017subdimensional,ma2018higher,pretko2017generalized,pretko2017finite,Ma2017-qq,Ma2017-cb,2018arXiv180302369Y,you2018symmetric}. These new quasiparticles, first seen in the context of exactly solvable spin models\cite{Haah2011-ny,Halasz2017-ov,Vijay2016-dr,Vijay2015-jj,Chamon2005-fc,hsieh2017fractons,Slagle2017-ne,devakul2018fractal,Halasz2017-ov,Hsieh2017-sc,shirley2018fractional,yoshida2013exotic,pai2018fractonic,prem2018pinch,slagle2018symmetric,2018arXiv180711479P,song2018twisted,gromov2018towards} only move within lower-dimensional manifolds such as planes, lines, or fractals. The subdimensional nature of fracton excitations gives rise to unconventional features including glassiness and subdiffusive dynamics \cite{Chamon2005-fc,prem2017glassy}. Besides, the restricted quasiparticle mobility makes fracton stabilizer codes promising for quantum-memory and quantum-computation applications \cite{kitaev2003fault,Haah2011-ny,Slagle2017-la,shirley2017fracton,prem2018pinch,pretko2017fracton,Pretko2017-ej,slagle2018symmetric,gromov2017fractional,2018arXiv180711479P,pai2018fractonic,2018arXiv180601855B,ma2018fracton,prem2017emergent,bulmash2018higgs,pretko2017subdimensional,ma2018higher,pretko2017generalized,pretko2017finite,Ma2017-qq,Ma2017-cb,bulmash2018braiding,bravyi2010majorana,Prem2017-ql,PhysRevB.94.085116}. 

While the theoretical study of fracton phase has been intensively focused on abelian fracton models, the non-Abelian side was less explored. In Ref.~\cite{vijay2017generalization}, the author introduces a $p+ip$ superconductor coupled to fracton gauge theory. Owing to the subdimensional nature of the gauge flux, the Majorana zero modes, trapped inside the $\pi$ flux (fracton gauge flux) have restricted mobility.
A further systematic investigation of non-Abelian fracton models was proposed in Ref.~\cite{huang2018cage} based on cage-net condensate.

In this work, we intend to design a non-Abelian particle in fracton matter from an alternative route. Instead of searching for non-Abelian fracton models, we start from an abelian fracton stabilizer code, known as the checkerboard model, and create non-Abelian properties via twisted defects. Albeit the abelian nature of the quasiparticle excitations in checkerboard model, the existence of the twisted defect, which permutes quasiparticle type, supports additional projective non-Abelian zero modes.

The twisted defect induced non-Abelian statistics can be traced back to the literature of topological order phase\cite{bombin2010topological,you2012synthetic,you2012projective,barkeshli2013classification,barkeshli2014symmetry,teo2013existence,barkeshli2012topological,barkTheoryofdefects,barktwist,kesselring2018boundaries,teo2016globally,khan2014anyonic,barkeshli2015generalized,tarantino2016symmetry,mesaros2013classification,PhysRevX.7.031048} from several parallel directions. In the mathematical physics side, an extrinsic defect permuting distinct quasiparticles can trap non-Abelian zero modes\cite{barkeshli2012topological,barkTheoryofdefects,barktwist,teo2016globally}. Based on these observations, a zoology of twist defects are proposed. The prominent examples are the Wen-plaquette models\cite{bombin2010topological}, where a lattice dislocation permutes $e$ and $m$ particle and hence supports projective Ising anyons. Such projective dislocation defect can be generalized to $Z_n$ quantum rotor model where a dislocation traps a $Z_n$ parafermion mode\cite{you2012projective,you2012synthetic}.  Besides, in a variety of quantum Hall bilayer systems, a Genon defect, which permutes the quasiparticle between layers\cite{barkeshli2012topological}, also supports non-Abelian zero modes whose quantum dimension is widely tunable by the K matrix of the bilayer quantum Hall system. 

Our construction paves new routes toward the realization of non-Abelian fracton phases. In particular, the twisted defect in checkerboard model can be `artificially' realized by imposing strong Zeeman field along a branch cut\cite{you2012synthetic}. Such synthetic twisted defect is highly amenable so one can move the defect to implement the braiding. 
The paper is organized as follow. In Section II, we review the parton construction of checkerboard model. In Section III, we demonstrate the non-Abelian nature of twisted defect in checkerboard model via Wilson loop deformation. In Section IV, we propose an experimental design for twisted defect in Majorana quantum Lego platform.

\section{Majorana construction for Fracton}

To search for non-Abelian defects in fracton phases, we first evoke the parton construction of fracton via Majoranas. As is pointed out in Ref.~\cite{youcode,you2018majorana,hsieh2017fractons}, most Fracton stabilizer codes have a Majorana Lego construction whose building block includes Coulomb blockaded Majorana islands and weak inter-island Majorana hybridizations. The spin Hamiltonian consisted of stabilizers terms has a mean-field description as a Majorana band model. Each site supports even number of Majorana fermions hybridized with nearby Majoranas. By imposing an onsite fermion parity constraint, the Majoranas are reduced to a spin(hyper-spin) degree of freedom whose interactions reproduce the fracton stabilizer codes.

\begin{figure}[h]
  \centering
      \includegraphics[width=0.4\textwidth]{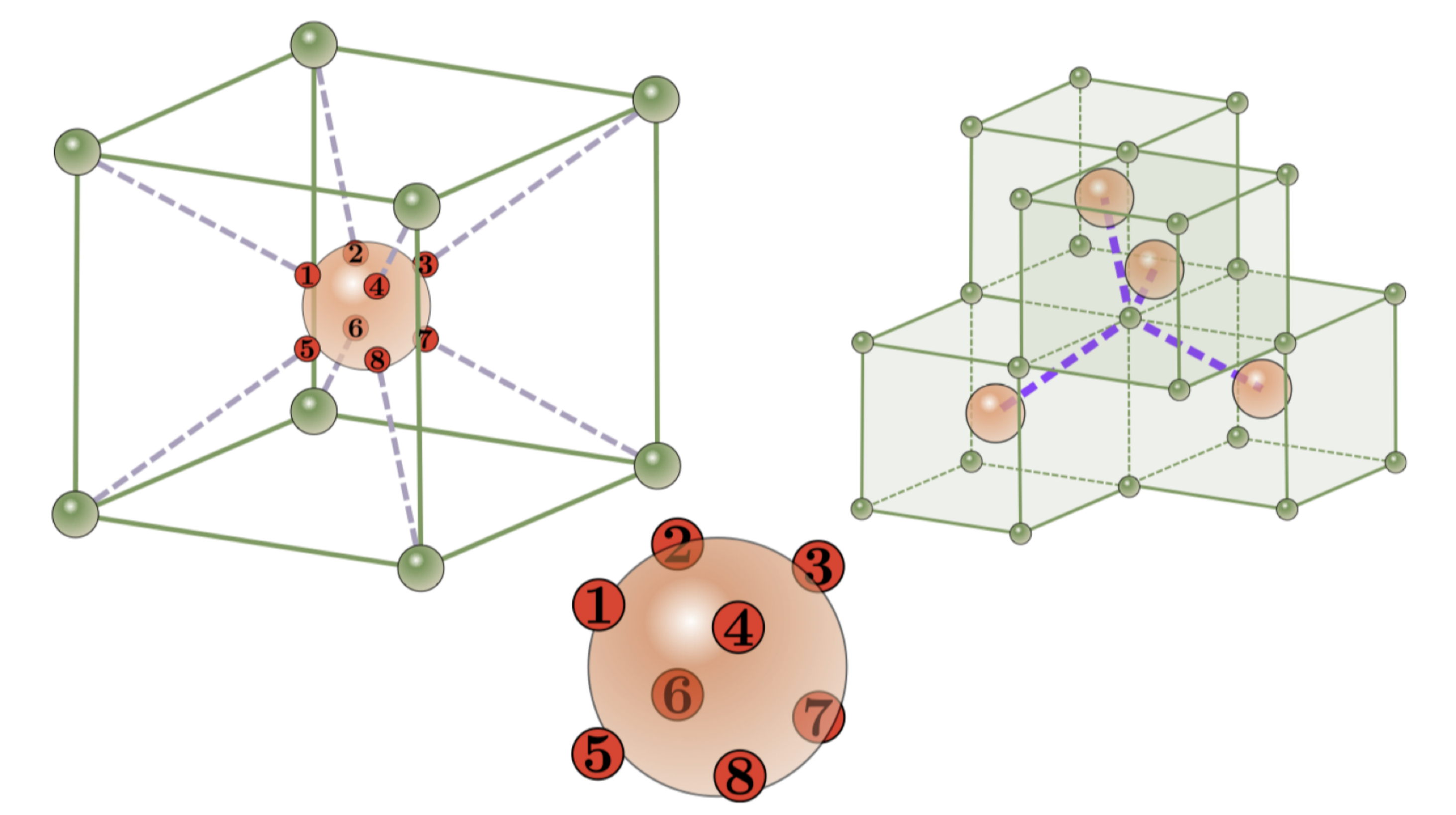}
  \caption{The red site in checkerboard cube center contains 8 Majorana hybridized with one of the 4 Majorana on green corner site. The dashed purple lines illustrate the hybridization between two Majoranas on nearby site.} 
  \label{levin4}
\end{figure}

To set the stage, we review the Majorana mean-field description of checkerboard model proposed in Ref.\cite{youcode}. The construction is based on a body-centered checkerboard lattice (Fig.~\ref{levin4}). Each green site at the vertices of the checkerboard lattice hosts four Majoranas (denoted by $\chi$). The red sites at the center of the cubes host eight Majoranas(denoted by $\eta$). The Majoranas on the green sites pair with the closest Majorana on the neighboring red site as shown in Fig.~\ref{levin4}.

\begin{figure}[t]
  \centering
      \includegraphics[width=0.4\textwidth]{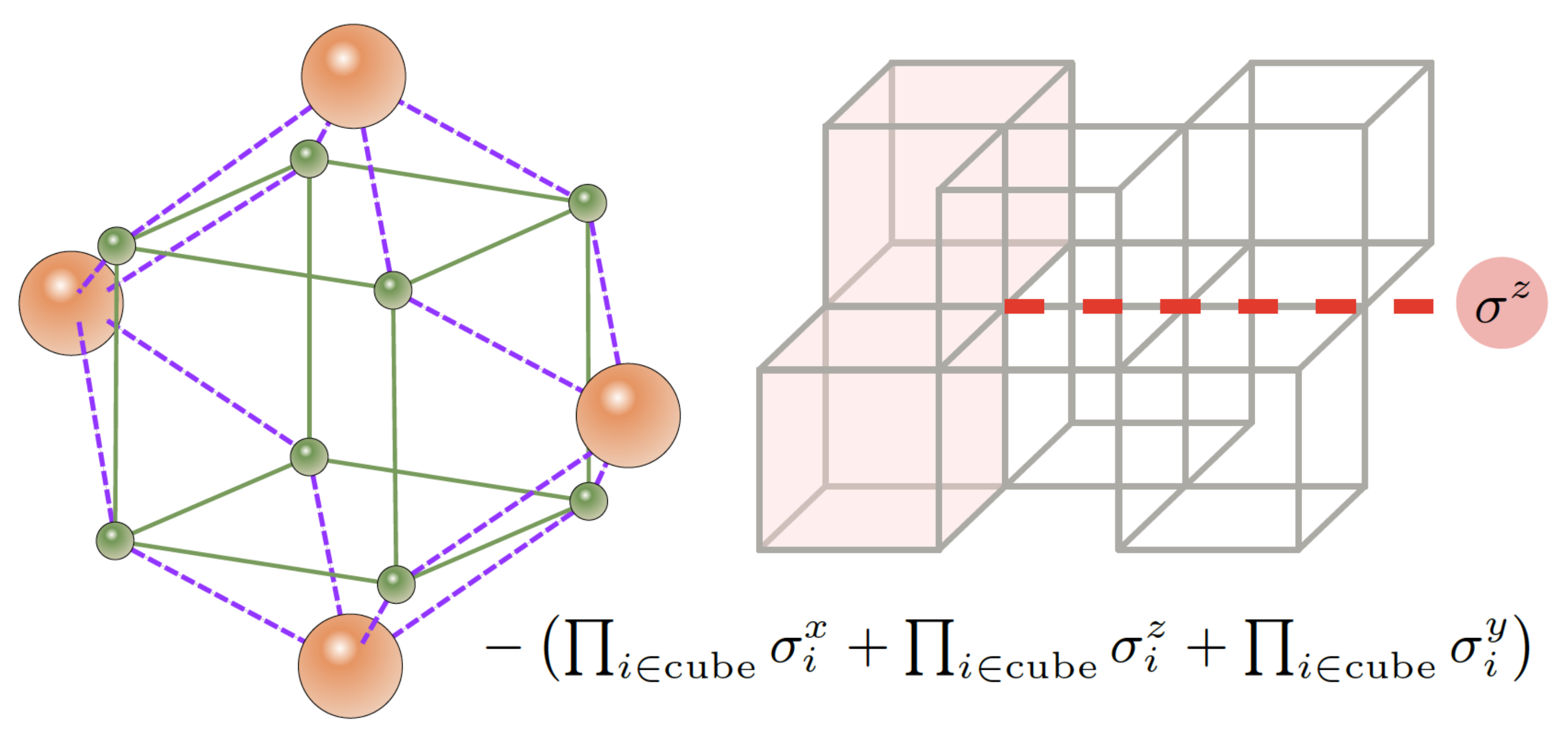}
  \caption{L: In strong U limit, the effective Hamiltonian consists the product of 16 Majorana pair(purple dashed line) surrounding the 4 side-faces of the cube. Each 16 Majorana pair can be expressed as the 8-spin cluster interaction on the checkerboard cube. R: The spin Hamiltonian displays fractonic excitation where a pair of cube-flip(red) excitation can only move along straight line by applying the $Z$ logical operator.} 
    \label{levin5}
\end{figure}

We now introduce the onsite interaction
\begin{align} 
&H_{\eta}=U(\eta_1 \eta_2 \eta_3 \eta_4+\eta_5 \eta_6 \eta_7 \eta_8
+\eta_1 \eta_4 \eta_8 \eta_5\nonumber\\
&\,\,\,\,\,\,\,\,\,\,\,\,\,\,+\eta_2 \eta_3 \eta_7 \eta_6+\eta_2 \eta_1 \eta_5 \eta_6+\eta_3 \eta_4 \eta_8 \eta_7)
\label{omg}
\end{align}
on each red site, corresponding to a sets of four-Majorana interactions on each face of the cube, together with a four-fermion interaction $H_{\chi}=U\chi_1 \chi_2 \chi_3\chi_4$ fixing the fermion parity for every green site. In the strong interaction limit, the latter projects the four Majoranas of a green site into a spin-${1}/{2}$ degree of freedom which we denote by $\sigma$. The interaction $H_{\eta}$ projects out any degree of freedom on the red site and it merely mediates interaction between green sites. As a result, the effective Hamiltonian in the strong interaction limit becomes
(see Fig.~\ref{levin5})
\begin{align} 
&H = - \sum_{\rm cubes} \left\{\prod_{j \in {\rm cube}}\sigma_j^x+\prod_{j \in {\rm cube}}\sigma_j^z + \prod_{j \in {\rm cube}}\sigma_j^y\right\},
\label{che}
\end{align}
which is known as the bosonic checkerboard model \cite{Vijay2015-jj,Vijay2016-dr,Vijay2017-ey}. 
The interactions involve products of eight spins on all checkerboard cubes without red sites at the center. The fundamental excitation of the checkerboard model, a single cube flip, is completely immobile, while pairs of cube flips on adjacent planes are restricted to move along a fixed direction. For a cube of linear dimension $L$ with periodic boundary conditions, the ground state degeneracy is $2^{6L-6}$ and different ground state sectors cannot be deformed into each other via local operators. 

\subsection{Twsited defect from onsite hybridization}

\begin{figure}[h]
  \centering
      \includegraphics[width=0.2\textwidth]{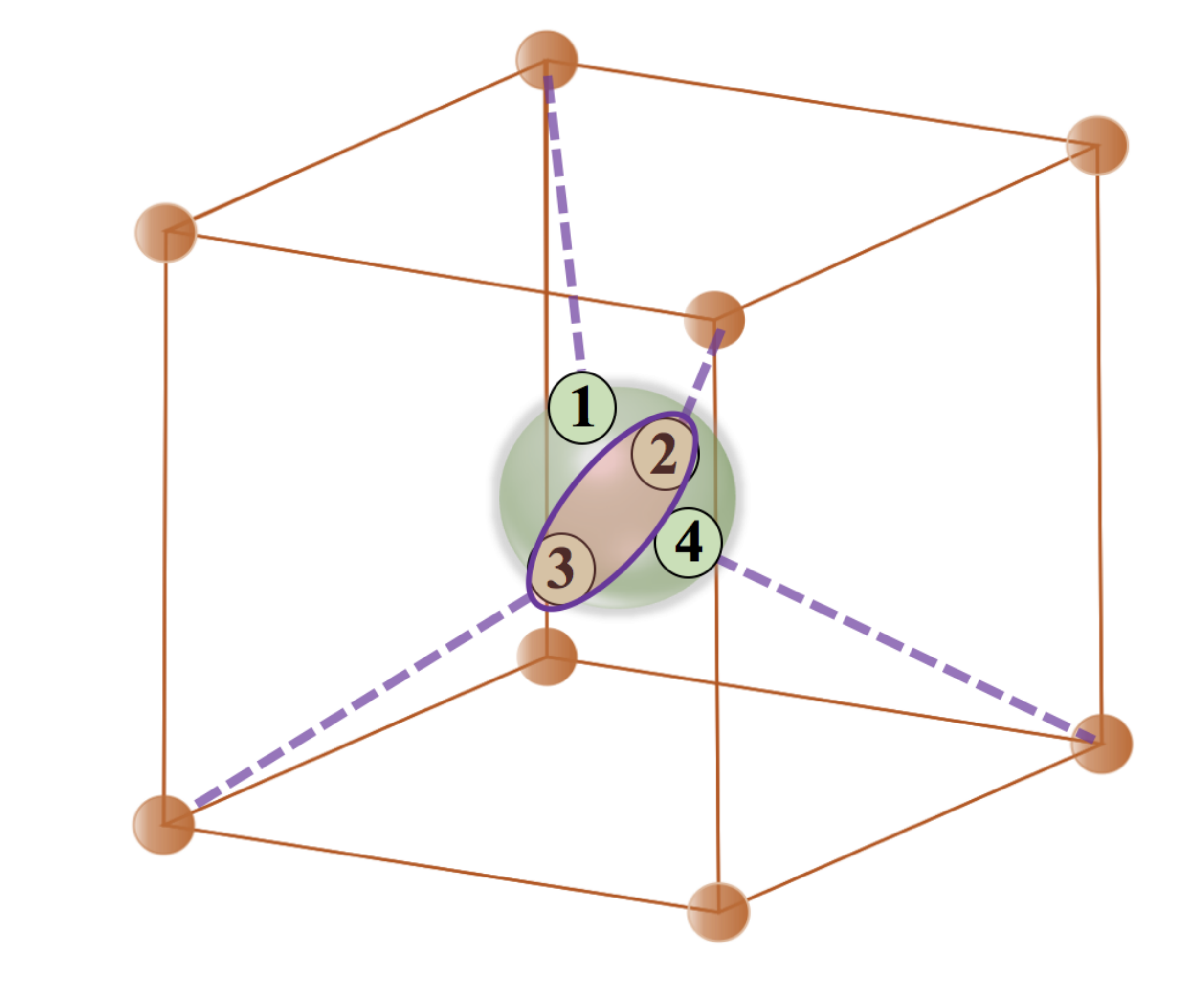}
  \caption{The red oval denotes a strong onsite hybridization between Majoranas.} 
    \label{onsite}
\end{figure}

\begin{figure}[h]
  \centering
      \includegraphics[width=0.38\textwidth]{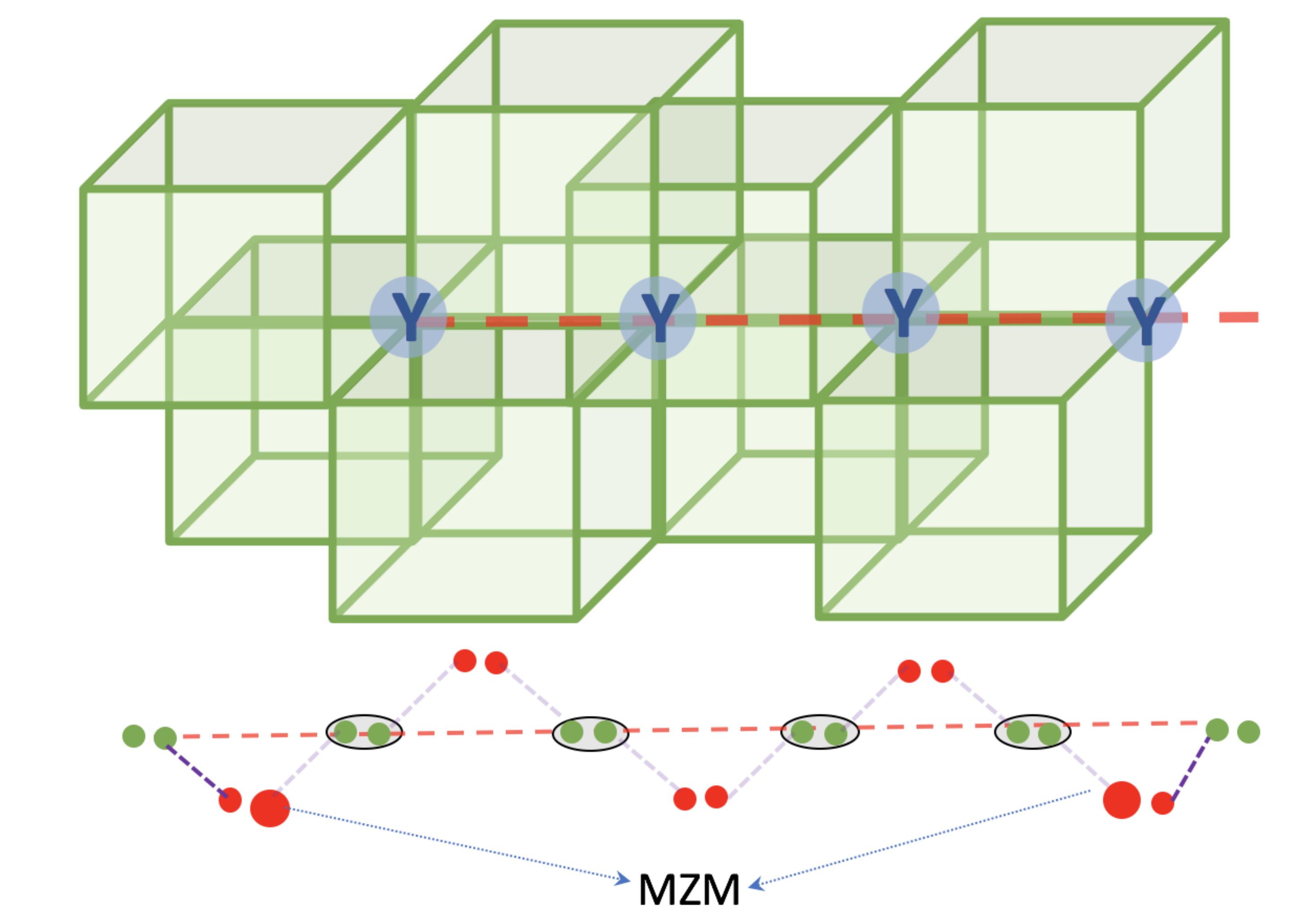}
  \caption{top: After various onsite projection, the onsite hybridization term on the green site creates a transverse Ising field $\sigma^y$ along the dashed red line.
  Bottom: The defect (red dashed) line resembles a Kitaev chain in the trivial phase with every two Majoranas hybridized within each site. Here red and green dot denote the Majorana on red or green sites. The grey oval denotes onsite hybridization and the enlarged red dot is the dangling MZM.} 
    \label{onsite2}
    \end{figure}

The parton construction in Ref.\cite{youcode} reveals a mean-field description of the checkerboard model as a topological superconductor with quasi-one-dimensional pairing structure. To engineer a synthetic twisted defect, we impose a strong onsite hybridization for the Majoranas on the green site along the defect line as Fig.~\ref{onsite}-\ref{onsite2} and leave the Hamiltonian invariant elsewhere. In the strong onsite hybridization limit, the defect line carries a 1$d$ p-wave superconductor in the trivial phase. At the end of the branch cut, there appears a kink between the topological and trivial superconductor, leaving an unpaired Majorana zero mode(MZM) as Fig.~\ref{onsite2}. As the dangling Majoranas living at the two ends of the branch cut are spatially separated, these zero modes are topologically robust against any perturbations.


\section{Checkerboard model with twisted defect} 

The parton construction, at the mean-field level, explicitly reveals the existence of MZM at the end of the defect line. But what happen if we impose the onsite projection in Eq.~\ref{omg} to the parton state? As the projection enforces even fermion parity constraint, any fermion zero mode carrying odd fermion parity should be projected out. To recognize the effect of twisted defect under fermion parity projection, we now scrutinize the checkerboard model in the presence of a defect line. 

\begin{figure}[h]
  \centering
      \includegraphics[width=0.4\textwidth]{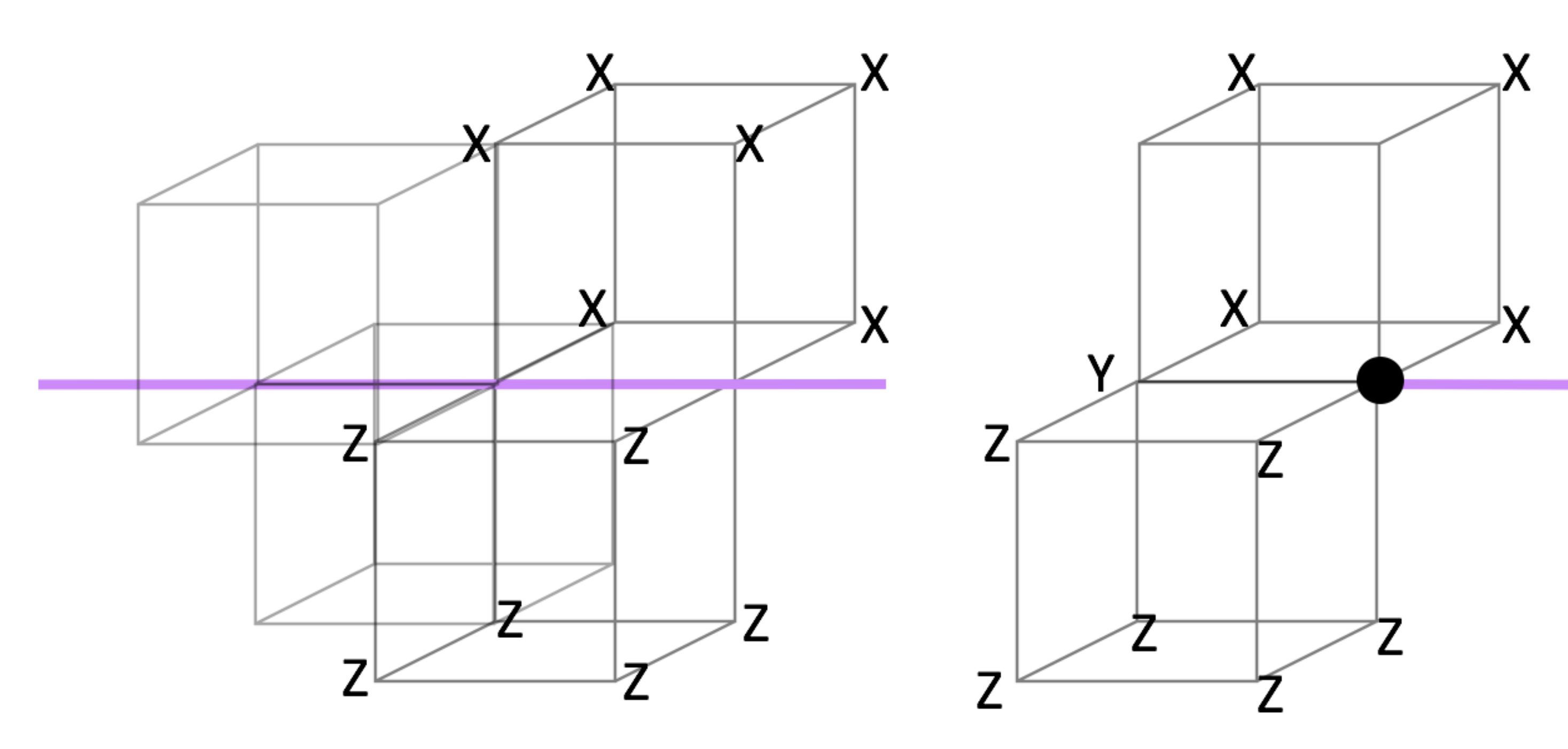}
  \caption{L: Stabilizer on the defect line. R: stabilizer at the end of defect line.} 
    \label{sta}
\end{figure}

After fermion parity projection, the onsite hybridization term creates a Zeeman field $J \sigma^y$ as Fig.~\ref{onsite2} on the green sites along the branch cut. In the strong Zeeman field limit, the X and Z cubic stabilizer in Eq.\ref{che} along the defect line is suppressed. The Zeeman field quenches the qubit degree of freedom along the branch cut as if they were removed from the lattice effectively. The remaining stabilizer between the defect line involves the cubic term with $Y$ stabilizer, as well as the product of two $X$ (or two $Z$) cubic stabilizer sharing a hinge on the defect line. Meanwhile, we also have terms from the product of $X$ and $Z$ cubic stabilizer sharing a hinge on the defect line which can be generated by the product of the previous ones. Fig.~\ref{sta} illustrates the stabilizer terms along the defect line(purple). Along the branch cut, each site experienced a strong Zeeman field $\sigma^y$. For the hinge-shared checkerboard cubes along the defect line, apart from the $Y$ stabilizer on each cube, there also exists a stabilizer with 12 logical operators which origins from the product of two cubic stabilizers overlapping a hinge on the branch cut. At the end of defect line, apart from the $Y$ stabilizer, there also exists a stabilizer with 13 logical operators which origins from the product of two cubic stabilizers at the end of the branch cut.

The strong Zeeman field condenses the quasiparticle generated by $\sigma^y$ operator along the branch cut. As a result, the $e$ type of quasiparticle generated by $\sigma^x$ string is permuted to the $m$ type of quasiparticle generated by $\sigma^z$ string.  In the meantime, the checkerboard model has a self-dual structure where both the Hamiltonian and quasiparticle type are invariant if we permute $\sigma^z$ and $\sigma^x$ terms. Such quasiparticle self-dual structure, known as the anyonic symmetry, is extensively studied in 2D topological order literature. In particular, a twisted defect which exchanges two abelian quasiparticles related by anyonic symmetry can support non-Abelian zero modes. One typical example of non-Abelian twisted defect is the Wen-plaquette model\cite{bombin2010topological}, where the $e$ and $m$ quasiparticles are related by translation symmetry. By developing a dislocation defect with odd Burgers vector, the dislocation line permutes $e$ string with $m$ string. Each dislocation contributes a $\sqrt{2}$ degeneracy which is exactly the quantum dimension of the Ising anyon. One can demonstrate the non-Abelian nature of dislocation defects via adiabatic modular transformations, or Dehn twists, of the topological state on the effective genus g-surface\cite{barkeshli2013classification,you2012synthetic}.

\subsection {Wilson loop deformation for defect braiding}
\label{braidxy}

We now generalize such non-Abelian twisted defect prototype to the checkerboard model.
To verify the non-Abelian nature of twisted defect, the first step is to check the unitary evolution of the ground state after exchanging two defects.
Assume the checkerboard model on the defect lattice supports a ground state manifold labelled by different topological sectors $\psi_a,\psi_b,...$. Initialize $\psi_0=\psi_a$, we exchange the two twisted defects adiabatically and reach the finally state $\psi_t$. With the condition that the final state becomes a coherent combination of several topological sectors in the ground state manifold $\psi_t=a\psi_a+b\psi_b+..$, the braiding process is non-Abelian. As the ground state manifolds of the checkerboard model with sub-extensive degeneracy are labelled by global flux configurations, the eigenvalue of the Wilson line operator characterize different topological sectors. Thus, any unitary evolution of the ground state under the braiding procedure is associate with the unitary evolution and deformation of the Wilson line operators.

In the checkerboard model, a single Wilson line operator, as illustrated in Fig.\ref{levin5}, is restricted to straight line configurations along $x,y,z$ direction. Any deformation of the Wilson line creates additional excitation at the corner and pulls the state out of the ground state manifold. Meanwhile, two adjacent Wilson lines parallel along $k$ direction can move on the $i-j$ plane\cite{Vijay2016-dr}. Any further deformation toward the $k$ direction would unavoidably create additional excitation.

\begin{figure}[h]
  \centering
      \includegraphics[width=0.3\textwidth]{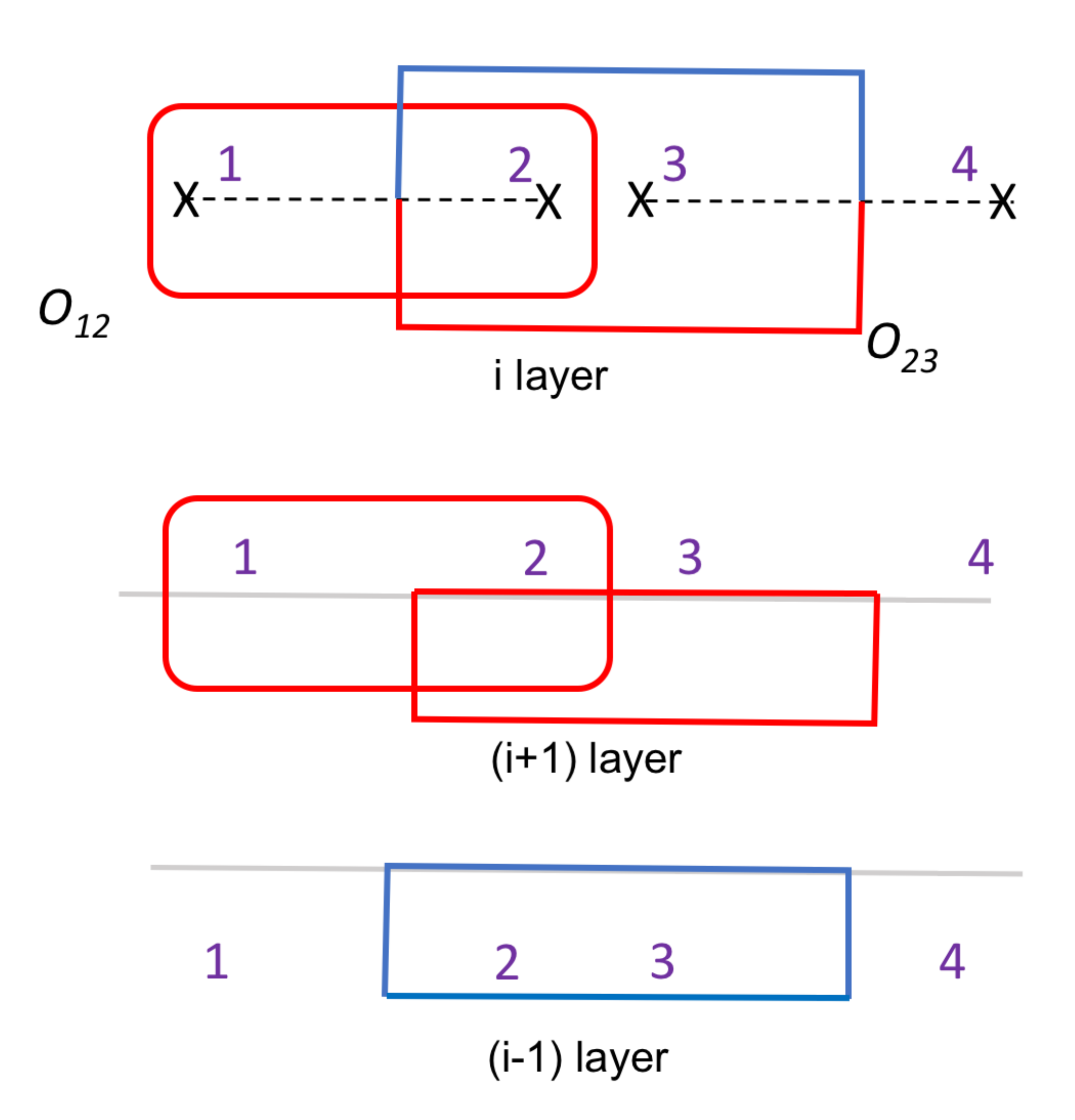}
  \caption{The Wilson line operator $O_{23},O_{12}$ defined on $i$ and $i\pm 1$ layer. Red and blue line denotes the X and Z operators. The twisted defect only exists on $i$ layer.} 
    \label{dis}
\end{figure}

In the presence of defect lines, we can define additional non-contractible Wilson line operators around the branch cut. As a single Wilson line is rigid and unbendable, we need to include two parallel Wilson lines from adjacent layers enclosing the branch cut as Fig.~\ref{dis}.

\begin{figure}[h]
  \centering
      \includegraphics[width=0.4\textwidth]{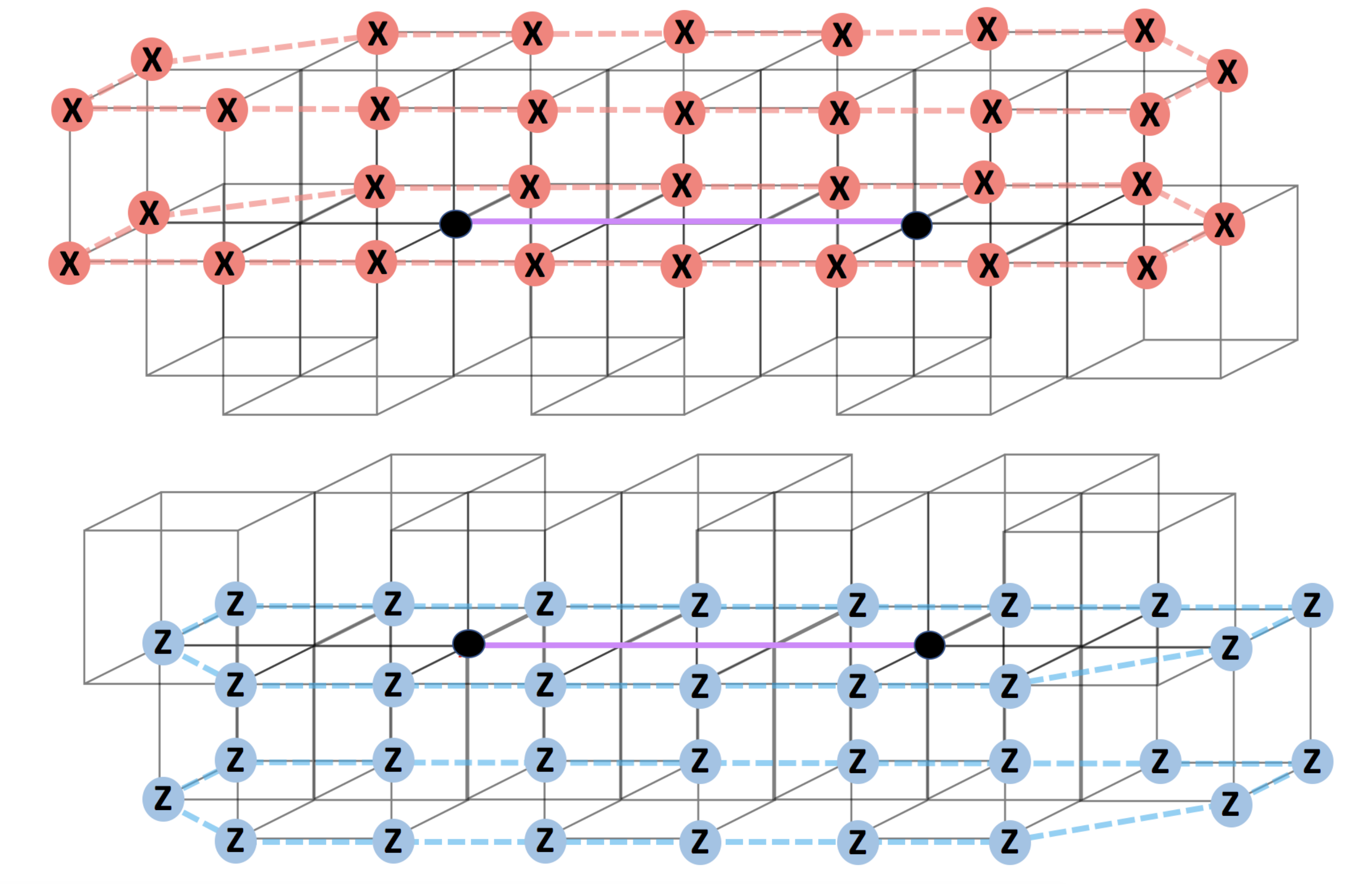}
  \caption{A detailed plot of the Wilson line operator $O_{12}$ enclosing the defect line. Such parallel Wilson line can be defined on $i$ and $i+1$ layer (top), or on $i$ and $i-1$ layer(bottom).} 
    \label{w1}
    \end{figure}

\begin{figure}[h]
  \centering
      \includegraphics[width=0.35\textwidth]{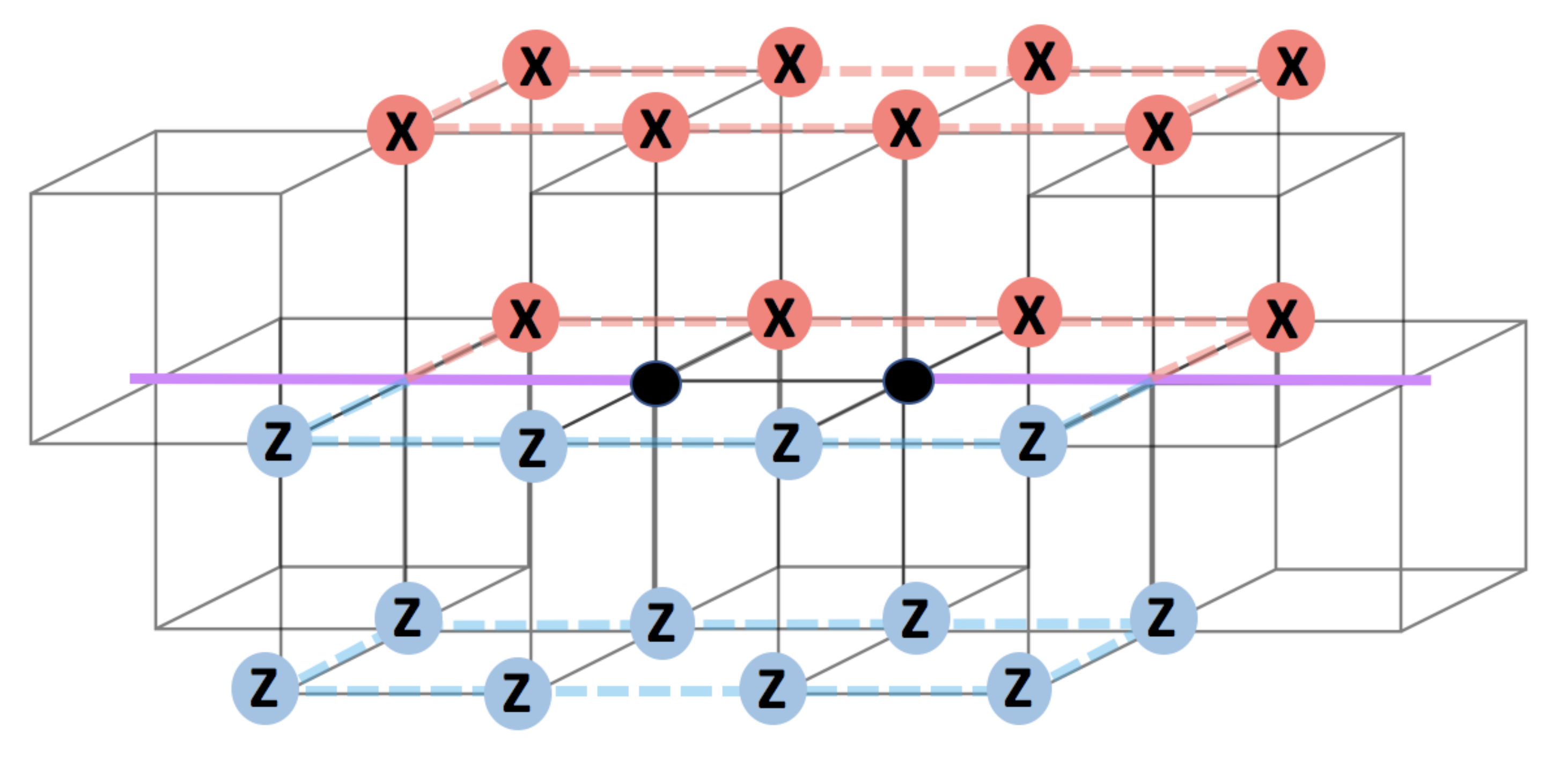}
  \caption{A detailed plot of the Wilson line operator $O_{23}$ enclosing around the defect line.} 
    \label{w2}
\end{figure}

Consider two defect lines living on the $i$ layer of the x-y plane arranged on the same row.
A non-contractible Wilson line operator $O_{12}$ requires two parallel loops with X operators on $i$ and $(i+1)$ layer as Fig.~\ref{w1}. The loop on the $i$ layer encircles the branch cut on the same layer. Despite the fact that the $i+1$ layer does not contain any branch cut, it is impossible to deform the loop on $i+1$ layer alone as such deformation creates additional excitation. As a single Wilson line is restricted to go straight along certain direction, the close Wilson loop surrounding the defect line must contain two parallel loops on adjacent layers. Another non-contractible Wilson line operator $O_{23}$ is generated by a Wilson line of X operator at the top semi-circle and Z operator at the bottom semi-circle connected at the branch cut on the $i$ layer as Fig.~\ref{w2}. Such operator, can be regarded as a close loop of X and Z operator living at the top and bottom half of the branch cut with mutual overlap along the defect line. As the spin on the defect line is polarized along $\sigma^y$ direction, the product of $\sigma^x \sigma^z$ is fixed to be a constant which can be dropped out. Meanwhile, there is a close Wilson loop of X on $(i+1)$ layer together with a close Wilson loop of Z on $(i-1)$ layer as Fig.~\ref{w2}. These two Wilson loops have fixed configurations parallel to the Wilson lines on $i$ layer and thus cancel the additional excitation created at the corner.

The Wilson line algebra $O_{12} O_{23}=-O_{23}O_{12}$ suggests there is at least a two-fold degeneracy created by the defect lines. We label the two-fold degeneracy as $|0 \rangle, |1 \rangle$,
\begin{align} 
O_{12}|0 \rangle= |0 \rangle, ~O_{12}|1 \rangle= -|1 \rangle,  O_{23}|0 \rangle=|1 \rangle, 
\end{align}

The braiding process can be handled by the deformation of non-contractible Wilson loops. Consequently, the unitary evolution of loop algebra specifies the unitary evolution of the ground state properties after braiding. To implement the twisted defect braiding, we apply the loop algebra approach developed in Ref.~\cite{you2012synthetic,barkeshli2013classification}. The main idea of loop algebra is to use non-contractible Wilson loops which label different topological sectors in the ground state manifold to keep track of the twist defect configurations in the system. The braiding effect between defects is represented by the algebra relation of the Wilson loops. As the ground state manifold is simply an irreducible representation space of the Wilson algebra, any nontrivial change in the Wilson loop operator alters the ground states specified by these loops, resulting in a statistical Berry phase during the braiding process.

\begin{figure}[h]
  \centering
      \includegraphics[width=0.4\textwidth]{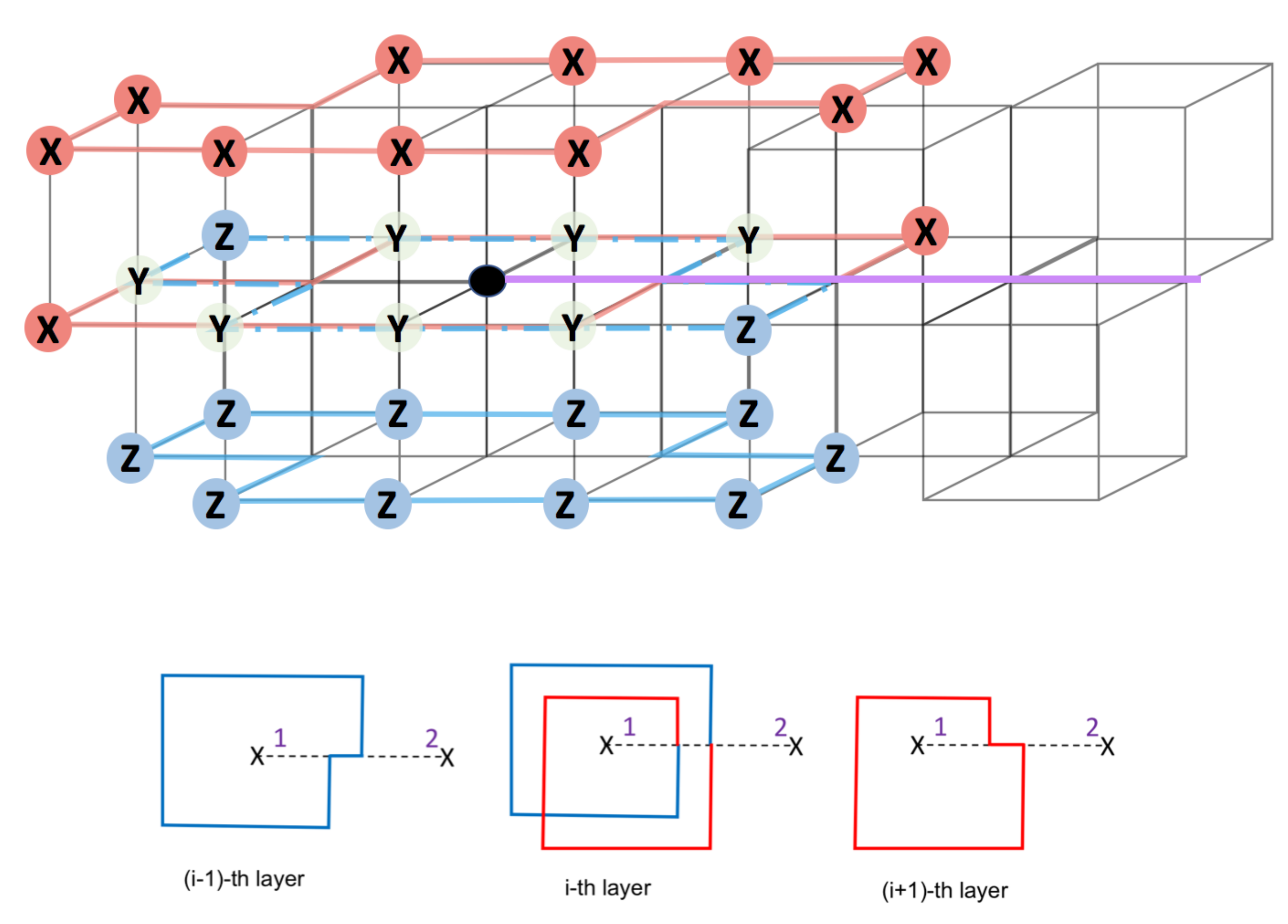}
  \caption{The purple solid line is the defect line. The red lines are X-operators with parallel loops on $i$ and $i+1$ plane. Meanwhile, the blue lines are Z-operators with parallel loops on $i$ and $i-1$ plane. } 
    \label{id}
    \end{figure}

Before we investigate the Wilson line deformation driven by braiding, we first introduce an identity operator enclosing the twisted defect. The identity operator incorporates the product of several stabilizers around the end of the defect line. Fig.~\ref{id} is a typical identity operator around the twisted defect. The red lines are X-operators with parallel loops on $i$ and $i+1$ plane. Meanwhile, the blue lines are Z-operators with parallel loops on $i$ and $i-1$ plane. On the $i$ plane, the X and Z operator always overlap odd times so there are odd number of Y operators. Such identity operator can be deformed provided the two loops on adjacent layers with the same color are parallel so there are no additional excitations created during the deformation. The local deformation of the identity operator corresponds to adding or removing stabilizers around the loops.

\begin{figure}[h]
  \centering
      \includegraphics[width=0.44\textwidth]{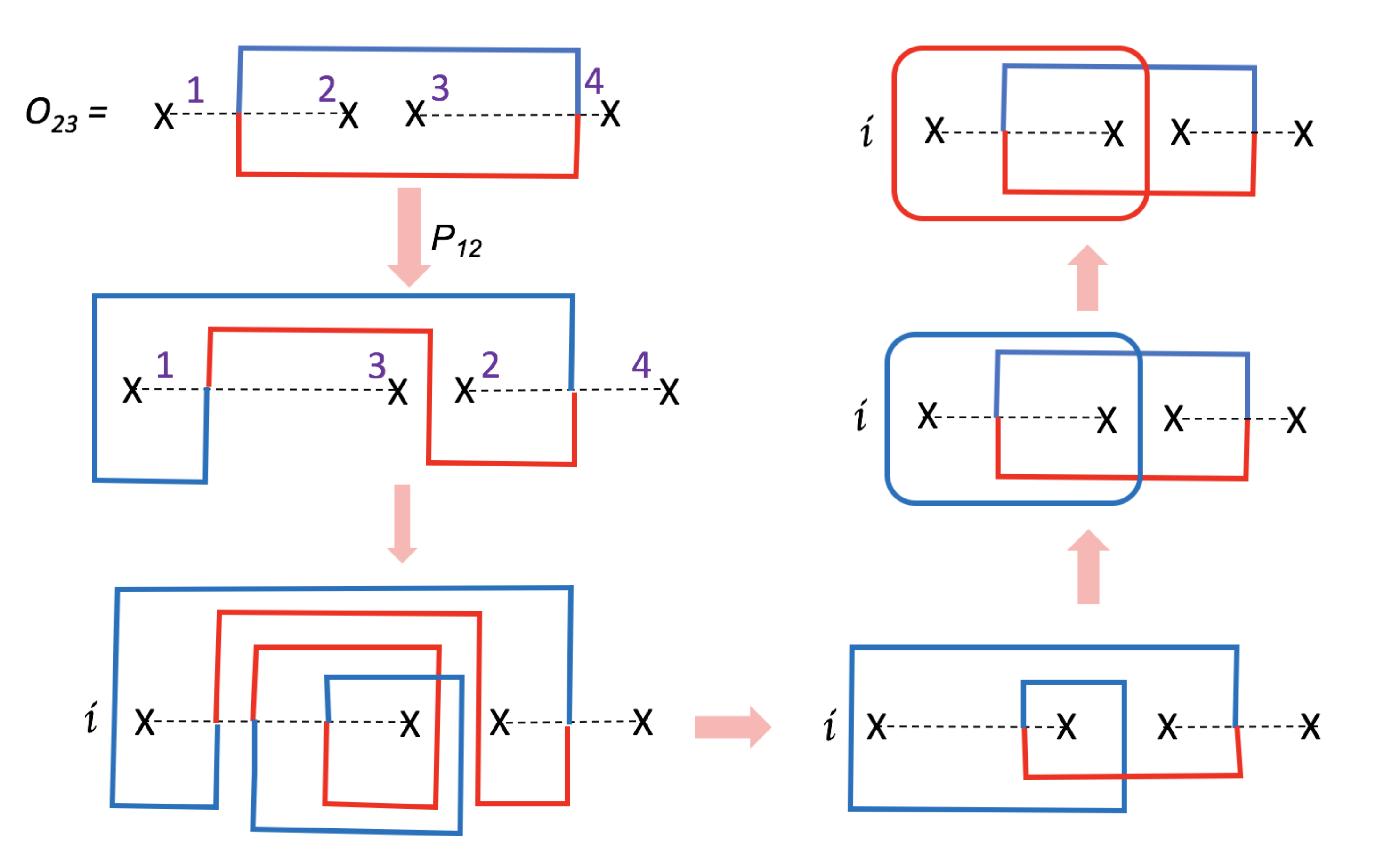}
  \caption{} 
    \label{deform1}
\end{figure}

Assume we have two pairs of twisted defects arranged along $x$ row. We choose two sets of non-contractible loops on the x-y planes as Fig.~\ref{dis}.
We first permute twisted defect 1 and 2. The Wilson line $O_{12}$ is obviously invariant under permutation. The deformation of $O_{23}$ is plotted as Fig.~\ref{deform1},

Based on the above graphic illustration, the braiding between defects is handled by the algebraic relations among these Wilson loop operators. The exchange between defects can be track by the deformation of Wilson line $O_{23}$. By inserting an identity operator around the defect 3, the Wilson line $O_{23}$ is finally deformed into the product of $ i O_{12} O_{23}$. Despite the graphic illustration in Fig.\ref{deform1} only plots the Wilson line at $i$ layer around the branch cuts, there always exist a Wilson line of $X$ parallel to the red loop on the above layer and a Wilson line of $Z$ parallel to the blue line on the beneath layer. These parallel loops are essential to ensure the Wilson line operators commute with the stabilizers in the Hamiltonian so the braiding process is adiabatic within the ground state. 

From this loop algebra approach, one can demonstrate that after exchanging defect 1 and 2, the Wilson line operators
transform as,
\begin{align} 
P_{12}:~O_{12} \rightarrow O_{12} , ~O_{23} \rightarrow i O_{12} O_{23}
\end{align}

\begin{figure}[h]
  \centering
      \includegraphics[width=0.46\textwidth]{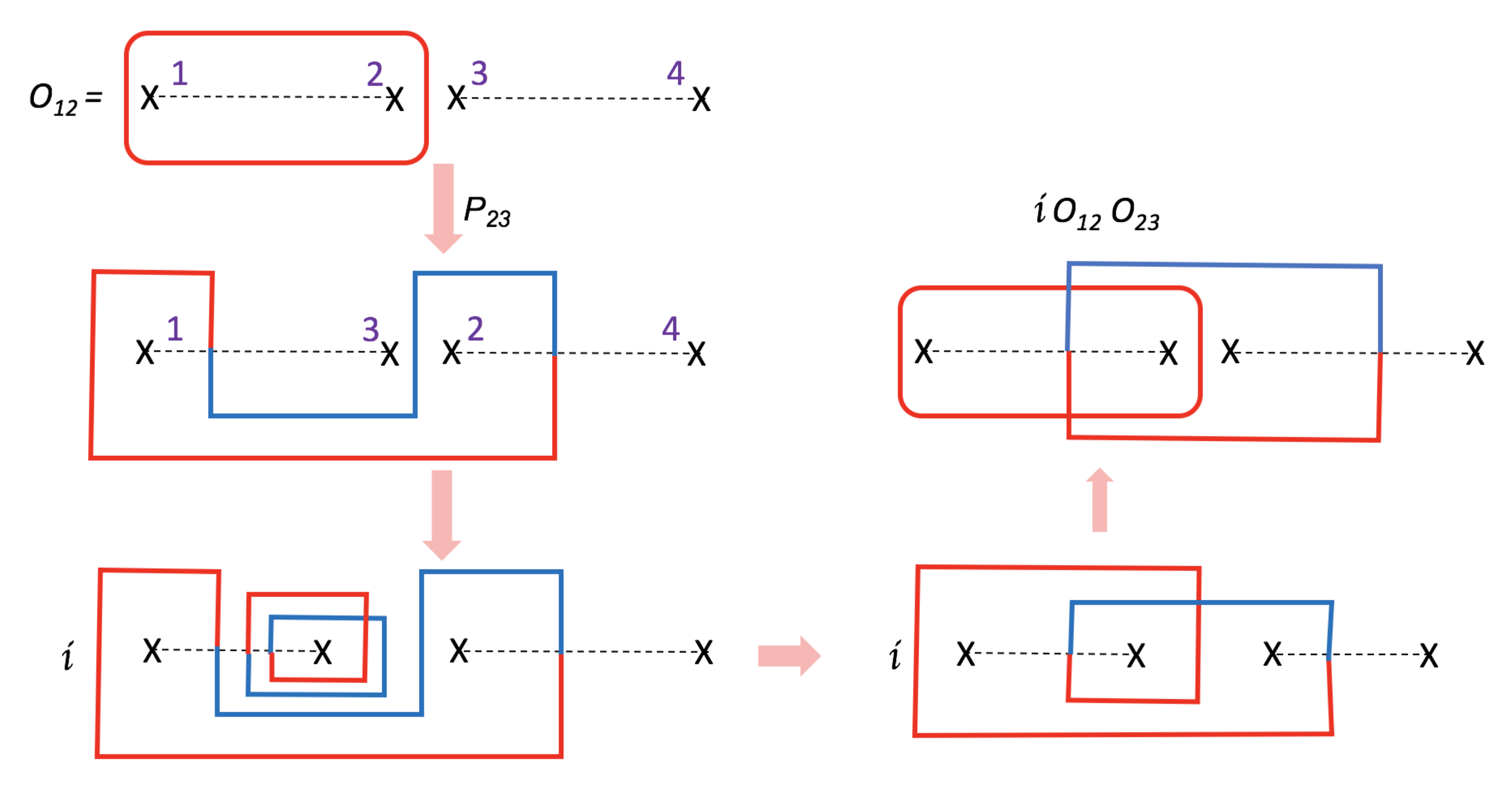}
  \caption{} 
    \label{deform2}
\end{figure}

One can utilize such loop algebra approach and permute between defect 3 and 2 as Fig.~\ref{deform2},

\begin{align} 
P_{23}:~O_{23} \rightarrow O_{23} , ~O_{12} \rightarrow iO_{23} O_{12}
\end{align}

As the Wilson line operators label different topological sectors in the ground state manifold, any deformation of the Wilson line operators indicates a unitary transformation of the topological sectors in the low energy Hilbert space.
One can figure out a set of unitary operators $T_{12}$ and $T_{23}$ satisfying
\begin{align} \label{TTT}
&T^*_{12}O_{12} T_{12} =O_{12} , ~T_{12}^* O_{23} T_{12} = i O_{12} O_{23},\nonumber\\
&T^*_{23}O_{23} T_{23} =O_{23} , ~T_{23}^* O_{12} T_{23} = i O_{23} O_{12},
\end{align}
One typical solution is,
\begin{align} 
& T_{12} =  |0 \rangle \langle 0 |+ i |1 \rangle \langle 1| \nonumber\\
& T_{23} = \frac{1-i}{2}(i|0 \rangle \langle 0 |+ i |1 \rangle \langle 1| +|1 \rangle \langle 0 |+ |0 \rangle \langle 1| )
\end{align}
There is a U(1) phase ambiguity as any phase shift $e^{i\theta_{12}}T_{12}$ still satisfies Eq.~\ref{TTT}. Hence, the twisted defects are `projective non-Abelian anyons' and the braiding group relation only holds up to a U(1) phase\cite{you2012projective,you2012synthetic,barkeshli2013classification}.
\begin{align} 
& T_{12} T_{23} T_{12} = e^{i\theta}T_{23} T_{12}T_{23} 
\end{align}

Finally, we would like to remark that the twisted defect, strictly speaking, are not `non-Abelian anyon' as they only appear as an extrinsic defects rather than a deconfined excitation. As the Berry phase accumulate from the defect exchange is topologically protected only up to a U(1) phase, these defects should be regarded as `non-Abelian zero modes' rather than `anyon with non-Abelian statistics'\cite{you2012projective,you2012synthetic,barkeshli2013classification}. 

\subsection{How to define `braiding of defects' in 3D fracton phases?}

During the braiding process discussed in Sec.~\ref{braidxy}, the two twisted defects, living on the same $x-y$ plane, are only exchanged within the $x-y$ plane. Such restricted motion on the 2D plane is essential for the non-Abelian Berry phase. Otherwise, two particles moving in full 3D space only display boson-fermion statistics.

For 3D fracton phases of matter, the quasiparticles have restricted mobility as a consequence of conservation laws. To move the fractonic quasiparticle out of the restricted direction, the creation of additional excitation is required accompanied by an energy barrier for such motion. As a result, the 2D anyon statistics is still well-defined between fracton particle excitations due to their restricted motion.

However, the twisted defect in our theory is an extrinsic defect rather than a deconfined excitation and its motion are not restricted as fractons. Nevertheless, the locations of the twisted defect do affect the ground state degeneracy and thus change the manifold of the low energy Hilbert space. If we have $m$ twisted defect lines along the same row, the degeneracy is $2^{6L-6+(m+2)}$. If we move one defect line out of the row but within the same plane, the degeneracy becomes $2^{6L-6+(m+3)}$. If the we move one defect line out of the plane, the degeneracy is $2^{6L-6+(m+4)}$. In checkerboard models, the cubic stabilizers on each plane product to unity so the total charge on each plane is conserved. As the defect line combines the cubic stabilizer terms along the branch cut, the charge conservation law on the plane around the branch cut no longer exists but the total charge conservation on adjacent planes still holds. Alternatively, one can view the defect line as the charge condensate along the branch cut which breaks the subsystem charge conservation on nearby planes.

When defining the Berry phase after unitary evolution in Eq.\ref{TTT}, we take the assumption that the system stays in the same ground state manifold. Based on this definition, the braiding procedure, or twisted defect exchange process shall keep the ground state manifold invariant. In Sec.~\ref{braidxy}, the deformation of the Wilson line operator implies that the ground state Hilbert space as an irreducible representation space of the Wilson algebra for $O_{12}, O_{23}$ remains unchanged. However, if we move one of the defect lines out of the x-y plane along the $z$ direction, the Wilson line operators $O_{12}, O_{23}$ in Fig.~\ref{dis} no longer exist and the ground state Hilbert space is changed accordingly. When we define the braiding process between defects on the x-y plane, there is an essential constraint that the ground state Hilbert space as an irreducible representation of the Wilson algebra for $O_{12}, O_{23}$ defined on the x-y plane remains unchanged. Likewise, we can always permute the twisted defect on the x-z plane(assume the branch cut is along $x$). The resultant braiding process is labelled by the Wilson line operators on the x-z plane and the whole exchange process keeps the irreducible representation space of such Wilson algebra invariant. To conclude, the braiding between twisted defects is well-defined only if we restricted the adiabatic process within a certain low energy Hilbert space, which turns out to be the irreducible representation space of the Wilson line algebra. Based on such definition, the defect braiding procedure requires restricted motion on a 2D plane so they obey 2D anyon statistics.

\section{experiment design}

The non-Abelian nature of the twisted defect in checkerboard model provides a promising avenue toward the realization of non-Abelian particles in three spatial dimensions. Motivated by the recent experiment proposal for realizing Fracton phases via Majorana quantum Lego\cite{you2018majorana}, we now propose an experimental design for engineering twisted defect in checkerboard models.

As is discussed in Ref.~\cite{you2018majorana}, the checkerboard model can be realized via Coulomb blockaded Majorana islands and weak inter-island Majorana hybridizations. The principal ingredients of such constructions are Majorana hybridization as well as local interactions which fix local fermion parities \cite{vijay2015majorana,2015arXiv150907135B,Landau2016SC,Sagi2018,Wille2018,Thomson2018,yang2018hierarchical}. These interactions is implemented by charging energy in Majorana islands, also referred to as Majorana Cooper pair boxes, which underlie current designs for Majorana-based topological qubits \cite{Lutchyn2018rev,vijay2015majorana,Karzig2017scale,Plugge2017box}. Each island contains even number of Majoranas, e.g., at the ends of semiconductor wires proximity coupled to a superconductor \cite{Lutchyn2010,Oreg2010,Lutchyn2018rev}. The island's charging energy fixes its fermion parity, corresponding to a multi-Majorana interaction.

\begin{figure}[h]
  \centering
    \includegraphics[width=0.4\textwidth]{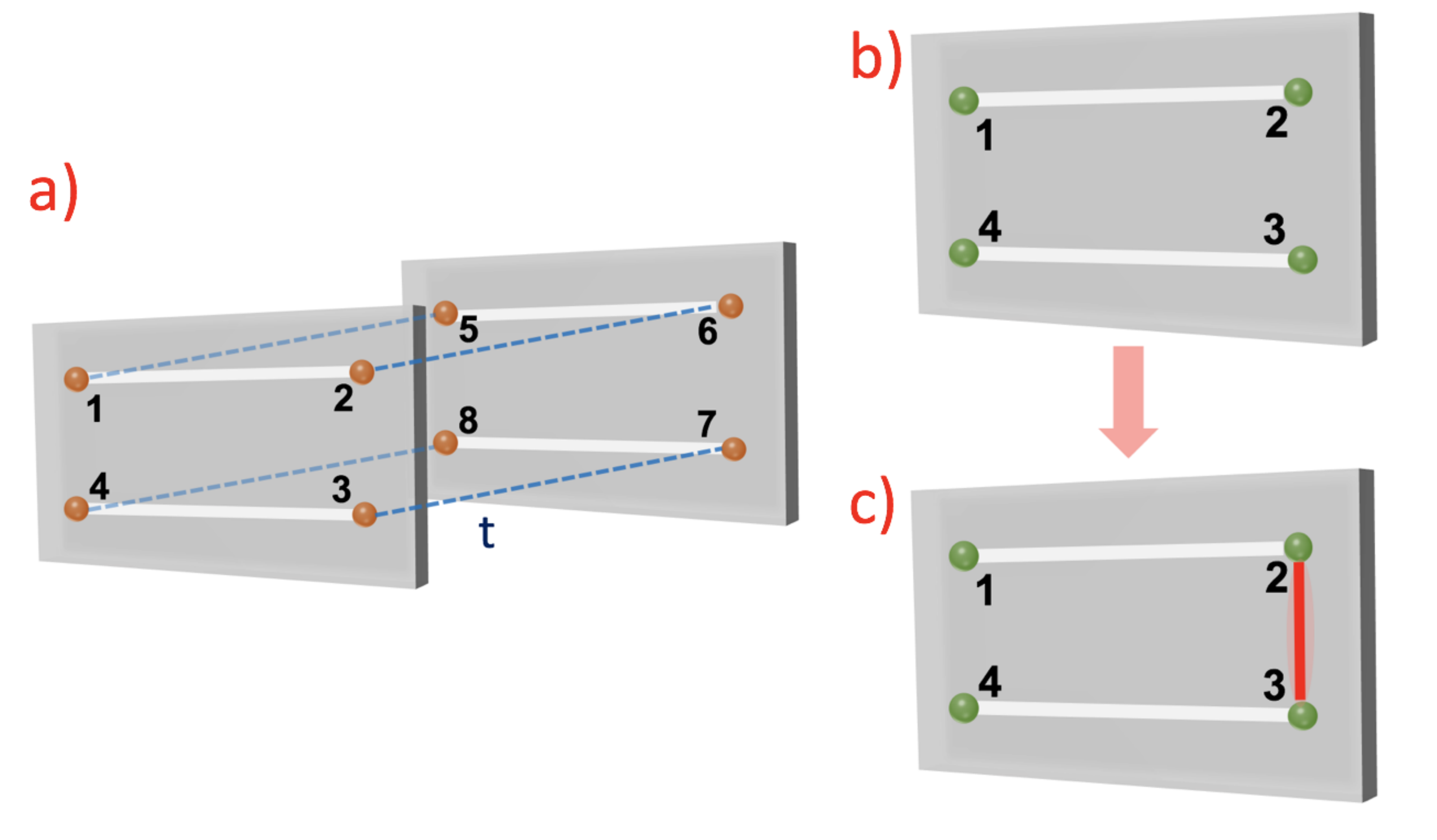} 
  \caption{a) There are two separate SCI (grey region) on each red site. The charging energy (U') fix the parity for $\eta_1 \eta_2 \eta_3 \eta_4$ and $\eta_5 \eta_6 \eta_7 \eta_8$ on each island. Tunneling between two SCI are denoted as the dashed blue lines. Such tunneling, in the strong U' limit, creates four Majorana interaction on the side faces. b) There is a single SCI with four Majoranas on each green site. The charging energy fix the parity. c) Onsite hybridization (red line) creates a Zeeman field $\sigma^y$. } 
    \label{exp}
\end{figure}

Based on the Majorana construction for checkerboard model in Ref.~\cite{youcode}, the green site living at the vertices of the checkerboard lattice hosts four Majoranas with even parity constraint.  We distribute the four Majoranas on each site onto a superconducting islands (SCI) as shown in Fig.~\ref{exp}. Each SCI is made from two semiconductor quantum wires proximity coupled to the same superconductor. The proximity-coupled quantum wires effectively realize open Kitaev chains with Majorana zero modes localized at their ends, so that there are a total of four Majoranas on each SCI. By virtue of their charging energy, each SCI can be tuned to have even fermion parity. 

The red sites at the center of the cubes each have eight Majoranas interact as Eq.~\ref{omg}.
To engineer this interaction, we first place the Majorana $\eta_1, \eta_2, \eta_3 ,\eta_4$ and $\eta_5 ,\eta_6, \eta_7, \eta_8$ on two separate SCI as shown in Fig.~\ref{exp}. The charging energies $U'(\eta_1 \eta_2 \eta_3 \eta_4+\eta_5 \eta_6 \eta_7 \eta_8)$ of these superconducting islands fix the parities $\eta_1 \eta_2 \eta_3 \eta_4$ and $\eta_5 \eta_6 \eta_7 \eta_8$. To generate the remaining four-Majorana interactions, we turn on inter-island hybridization 
\begin{align} 
&H_{t'}=it'(\eta_1 \eta_5+\eta_2 \eta_6+\eta_3 \eta_7+\eta_4 \eta_8),
\end{align}
where $t'$ controls the hybridization. In the low-energy limit with fixed parities for the two islands, one then obtains the effective Hamiltonian
\begin{align} 
&H_{\rm eff}=U'(\eta_1 \eta_2 \eta_3 \eta_4+\eta_5 \eta_6 \eta_7 \eta_8)\nonumber\\
&+O(\frac{t'^2}{U'})(\eta_1 \eta_4 \eta_8 \eta_5+\eta_2 \eta_3 \eta_7 \eta_6).
\end{align}
This exactly reproduce the interaction in Eq.~\ref{omg}. Importantly, the anisotropy of the coefficients does not affect the ground state manifold. In the regime $U'\gg t'\gg t$, the interaction projects out all the degree of freedom on red sites. 

To engineer a twisted defect with strong Zeeman field along the branch cut, we add intra-island Majorana hybridizations $i \chi_2 \chi_3$  for all SCI on the defect line as Fig.~\ref{exp}. After parity fixing via charging energy, the hybridization term becomes the transverse field $\sigma^y$ for the green site.
Such intra-island Majorana hybridizations can in principle be implemented by direct tunnel coupling. To enhance the onsite hybridization and create a strong Zeeman field along the defect, one can bridge two Majorana islands via a coherent link \cite{Karzig2017scale} consisting of an additional proximity-coupled quantum wire with its fermion parity fixed by charging energy. Its two Majorana end states would then be tunnel coupled to the two Majoranas of the Majorana islands which one wants to hybridize. Since the hybridization between the Majoranas on the coherent link and the islands can be realized through gate controlled tunnel junctions, the strength of the Zeeman field is highly tunable. 

The tunability of twisted defect from Majorana hybridization allows us to manipulate the position of the twisted defect and thus execute the braiding process directly. Apart from that, one can also apply the quantum non-demolition measurement scheme proposed in Ref.~\cite{zheng2015demonstrating} based on Majorana parity measurement to demonstrate the non-Abelian nature of the twisted defect. As the checkerboard model with twisted defect is built based on Majorana Lego, all spin operators are represented by Majorana parity operators. The measurement for spin correlation function is fulfilled by coupling several Majoranas to a single-level quantum dot. The resulting energy shift of the quantum dot level will then depend on the Majorana parity. 

\section{Conclusion and Outlook}
In the present work, we have introduced the notion of twisted defect in abelian fracton phase of matter. Such defect permutes quasiparticle types and further supports non-Abelian zero modes. In particular, we proposed an experiment-amenable construction of twisted defect by applying onsite Majorana hybridization or external Zeeman field to the branch cut line. Our result opens a new avenue for the search of non-Abelian quasiparticles in fracton phases. As the checkerboard model is a CSS code, one can generalize the current checkerboard model to its $Z_N$ counterpart. The twisted defect in such $Z_N$ checkerboard model, living at the end of the $e-m$ branch cut, supports parafermion zero mode\cite{you2012projective}. 

Motivated by the current result, several open questions arose, such as 1)~What type of twisted defects in fracton models support non-Abelian zero mode? Do such non-Abelian defects exist in type-II fracton models? 2)~ In principle, a general twisted defect is synthesized by anyon condensation along the defect line(membrane). Is there a systematical way to study the twisted defect effect via anyon condensate in fracton models?  3) In our current literature, the twisted defect permutes two distinct quasiparticles with the same mobility. What happens if a twisted defect permutes quasiparticles with different subdimensional mobility? 4) Recent studies of anyon condensate and gappable boundary\cite{youcode,bulmash2018braiding} in fracton phase of matter suggest the existence of zero modes on the hinge between two boundaries with distinct anyon condensate. Is there a comprehensive relation between twisted defect and distinct gapped boundary conditions? We raise these questions and aspects as the closing remark of this paper and leave them for future investigations.

\begin{acknowledgments}
We are grateful to Yi-Zhuang You for insightful comments and discussions as well as Daniel Litinski for preparing part of the figures.
YY is supported by PCTS Fellowship at Princeton University. 
\end{acknowledgments}

 \end{document}